\documentclass[conference, letterpaper]{IEEEtran}

\ifCLASSOPTIONcompsoc
  \usepackage[nocompress]{cite}
\else
  \usepackage{cite}
\fi

\ifCLASSINFOpdf
\else
\fi
\usepackage{multirow}
\usepackage{graphicx}
\usepackage{amsfonts}
\usepackage{amsmath}
\usepackage{mathtools}
\usepackage{makecell}
\usepackage{enumitem}
\usepackage{threeparttable}
\usepackage[caption=false]{subfig}
\usepackage{enumerate}

\newcommand{\n}{PRID}

\begin{document}
\title{
Tackling Multipath and Biased Training Data for\\
IMU-Assisted BLE Proximity Detection
}

\author{
\IEEEauthorblockN{Tianlang He${}^1$, Jiajie Tan${}^1$, Weipeng Zhuo${}^1$, Maximilian Printz${}^2$, S.-H. Gary Chan${}^1$ }
\IEEEauthorblockA{Department of Computer Science and Engineering\\
The Hong Kong University of Science and Technology, Hong Kong, China\\
Email: ${}^1$\{theaf, jtanad, wzhuo, gchan\}@cse.ust.hk, ${}^2$mprintz@connect.ust.hk}}


%


\maketitle

\footnotetext{This work was supported, in part, by Hong Kong General Research Fund~(under grant number 16200120).}

\begin{abstract}
  Proximity detection is to determine whether an IoT receiver is within a certain distance from a signal transmitter. Due to its low cost and high popularity,   Bluetooth low energy~(BLE) has been used to detect proximity based on the received signal strength indicator~(RSSI). To address the fact that RSSI can be markedly influenced by device carriage states, previous works have incorporated RSSI with inertial measurement unit~(IMU) using deep learning. However, they have not sufficiently accounted for the impact of
  multipath. Furthermore, due to the special setup, the IMU data collected in the training process may be biased, which hampers the system’s robustness and generalizability. This issue has not been studied before.

We propose PRID, an IMU-assisted BLE {\bf p}roximity detection approach  robust against {\bf R}SSI fluctuation and {\bf I}MU  {\bf d}ata bias. PRID histogramizes RSSI to extract multipath features and uses carriage state regularization to mitigate overfitting due to  IMU data bias. We further propose PRID-lite based on a binarized neural network to substantially cut memory requirements for resource-constrained devices. We have conducted extensive experiments under different multipath environments, data bias levels, and a crowdsourced dataset. Our results show that PRID significantly reduces false detection cases compared with the existing arts~(by over 50\%). PRID-lite further reduces over 90\% PRID model size and extends 60\% battery life, with a minor compromise in accuracy~(7\%).

\end{abstract}


\IEEEpeerreviewmaketitle

\section{Introduction}
\label{sec:intro}
Proximity detection is to decide whether
an IoT receiver is within a certain distance, say 5 meters, from a signal transmitter.  
If so it is a ``proximity'' event, or ``no proxmity'' event otherwise.
Free from human operation,
such decision  enables many 
proximity-based services (PBS),
such as contact tracing~\cite{ngCOVID19YourSmartphone2020,li_spatial-temporal_2021,TraceTogether,hatkeUsingBluetoothLow2020}, proximity marketing~\cite{zaimBluetoothLowEnergy2016}, and presence or check-in/checkout logging~\cite{yilangwuCICOSystemBased2015,caoDistanceEstimationMethods2018}.  These services often
need to detect proximity  independent of receiver carriage state, i.e., independent of whether the receiver is held in swinging hands, read positions, a pocket, backpack, side bag, and so on.

Many signals have been studied for proximity detection, such as radio frequency~(RF)~\cite{9496688,SEC14}, ultrasound~\cite{yoonEfficientDistanceEstimation2018}, and LiDAR~\cite{ramasamy2016lidar}.
Among them, Bluetooth Low Energy~(BLE) emerges as the most promising due to its low cost, low power consumption,
appropriate coverage range~(around 10 meters),
and wide availability in IoT devices.
In this work, we consider a common BLE-based PBS deployment scenario
where no hard partition~(or wall) cuts between
the receiver and the transmitter at the time of proximity detection.~(Detecting partition between two devices is
an independent issue outside the scope of this study.  Interested readers may refer to~\cite{xiaoNonLineofSightIdentificationMitigation2015,zhouLiFiLineOfSightIdentification2014} and references therein.)

Traditionally, 
proximity is detected by measuring the received signal strength indicator~(RSSI) and correlating it with 
distance, with the 
intuition that a lower RSSI means larger distance, and vice versa.
However, RSSI may be severely affected by
environment and receiver carriage state, leading to signal fading, fluctuation, and attenuation.
To address that, much of the existing work
employs deep learning to find a high-dimensional classification boundary from the RSSI values~(see, for example, \cite{shankarProximitySensingModeling2020a,chandelProxiTrakRobustSolution2020}).
However, these models are often not robust against RSSI fluctuation because the received signals may be
randomly affected by nearby mobile objects~(e.g., pedestrians, vehicles, rotating fans, etc.).
As a result, the decision
may be noisy, leading to less than satisfactory results.

To account for the impact of carriage state on RSSI, inertial measurement unit (IMU) readings are often collected as training data~\cite{shankarProximitySensingModeling2020a}.
Unfortunately, such a collection process does not always  faithfully reflect the operating condition in reality~(i.e., independence of carriage state in the general case).  In other words,
the training data may be biased, i.e., 
the dataset does not guarantee the carriage state, as given by IMU readings, to be independent of proximity.
%
If such a bias is not accounted for properly,
deep learning models could correlate the IMU readings to proximity, leading to  overfitting and generalization issues, and
performance 
highly dependent on the data acquisition process.

\begin{figure*}
\centering
\includegraphics[width=0.9\textwidth]{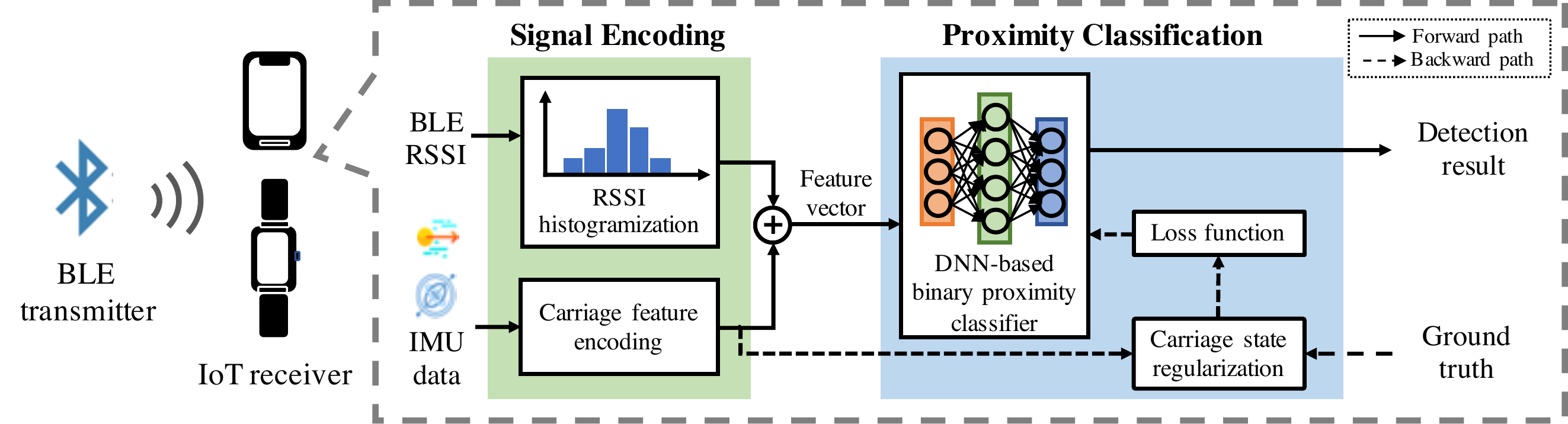}
\caption{
System overview of \n{}.
}
\label{fig:system}
\end{figure*}

To tackle the above multipath and biased training data issues, 
we propose \n{}, a novel, accurate, and generalizable IMU-assisted BLE {\bf p}roximity detection which is robust against {\bf R}SSI fluctuation and {\bf I}MU {\bf d}ata bias.
We overview \n{} in Figure~\ref{fig:system}.
In the forward (online) path,
\n{}, using a sliding window~(a few seconds), encodes BLE RSSIs and IMU data as feature vectors using RSSI histogramization and carriage feature encoding modules, respectively (signal encoding step).
After concatenating the feature vectors,
a trained binary classifier based on deep neural network~(DNN) is employed to detect proximity (proximity classification step).
In the backward~ path for offline training, \n{} employs carriage state regularization to reduce the IMU overfitting problem of the classifier.


The major contributions of this work are as follows:
\begin{itemize}
    \item \emph{RSSI histogramization to mitigate multipath impact}:
Multipath leads to a spread in RSSI distribution, which sheds lights on the features in the environment.
Therefore, to mitigate RSSI fluctuations due to multipath, we propose to represent the RSSI time series within a sliding window as a histogram. Through such a ``histogramization'' process, the multipath environment is modeled as a distribution for training and inference purposes.

    
    \item \emph{Carriage state regularization for potentially biased training data}:
We study, for the first time, how to mitigate
IMU training data bias  for BLE proximity detection.
We propose carriage state regularization, which first employs importance sampling
to reduce the discrepancy between the IMU features of the ``proximity'' state and that of the ``no proximity'' state in the training data.
Then, it applies the resultant sampling weights to a loss function in the training step, thereof effectively reducing IMU overfitting in the DNN-based proximity classifier.

\item \emph{\n{}-lite: Achieving memory efficiency for resource-constrained IoT devices:}
While
\n{} is efficient and deployable on smartphones commonly available on the market nowadays,
its memory requirement may still be demanding for some low-end IoT devices with extremely constrained resources, such as contact tracing tokens~\cite{TraceTogether}, smart car keys~\cite{caoDistanceEstimationMethods2018}, and audio guide devices for museums~\cite{jimenezFindingObjectsUsing2017a}. 
To further extend its use to resource-constrained IoT devices,
we propose \n{}-lite, a lightweight variant of \n{} achieving high memory efficiency with little cost in performance.
We employ model binarization on the DNN-based proximity classifier to
achieve a proper trade-off between
neuron quantity and neuron precision.
Moreover, due to the memory-efficient bit-wise operation between binarized neurons, \n{}-lite is more energy-conserving than floating-point manipulation and computationally efficient.

\end{itemize}

\n{} and \n{}-lite are easily implementable and  deployable.
We have developed them on commercial smartphones and IoT devices (Android smartwatch, Raspberry Pi Zero, and ESP-32).
To validate our design,
we  conduct extensive experiments on multiple sites with different multipath environments and levels of IMU data bias, including
the TC4TL challenge dataset~\cite{tc4tlChallenge2020}, which is a crowdsourced public dataset for automated contact tracing.
Our results show that \n{} achieves  substantial improvement as compared with the state-of-the-arts (with more than 50\% reduction in false detection cases in our dataset). 
 \n{}-lite reduces the model size of \n{} by 94\% (from 5MB to 0.3MB) and extends battery life by more than 60\% in our implementations, with only a minor compromise in accuracy (7\% reduction in F-score).  This demonstrates that \n{}-lite is deployable in most resource-constrained devices. 

The remainder of this paper is organized as follows. We discuss related work in Section~\ref{sec:related}, and how \n{} encodes RSSI and IMU signals in Section~\ref{sec:system_feature}. In Section~\ref{sec:system_classification}, we present \n{}'s online proximity classification and its offline training with carriage state regularization. In Section~\ref{sec:PRID-Lite}, we detail \n{}-lite. Later, we cover illustrative experimental results in Section~\ref{sec:exp} and conclude in Section~\ref{sec:conclude}.

\section{Related Work}
\label{sec:related}
Distance, multipath environment, and carriage state are three major factors that affect RSSI and hence proximity detection. In the following, we discuss previous works on these three factors.

Most of the early works on BLE proximity detection only consider the relationship between RSSI and distance. 
By assuming BLE signal merely attenuates over distance, they investigate how RSSI decreases as distance goes up by regression approaches -- 
it is intuitive to determine proximity events by ranging from RSSI.
The most common regression model is the log-distance path loss model~\cite{ercegEmpiricallyBasedPath1999}, which builds the exponential relationship between RSSI and distance. Some of its variants can be found in~\cite{phillipsSurveyWirelessPath2013}. 
Linear and inverse proportion relations are also studied in~\cite{alqathradyImprovingBLEDistance2017}. 
However, since these works fail to consider the multipath environment and carriage state, such regression approaches are mostly unreliable in practice.

Many recent works have considered the RSSI distortion caused by the multipath environments.
Their approaches are either based on noise reduction or sequential correlation upon an RSSI sliding window. 
The noise reduction works consider the RSSI distortion as signal noise 
and study various signal filters against multipath environments.
Works in~\cite{alqathradyImprovingBLEDistance2017, montanariStudyBluetoothLow2017,marateaNonParametricRobust2018, chandelProxiTrakRobustSolution2020} leverage statistical features (such as mean and median) for noise reduction. 
Other works in~\cite{zafariEnhancingAccuracyIBeacons2017,lamImprovedDistanceEstimation2018,pallaviOverviewPracticalAttacks2019,mackeyImprovingBLEBeacon2020} apply Bayesian filters to process RSSI.
Although these works improve detection accuracy in the settings under study, 
they suffer from poor extensibility to different environments since they assume the identical noise distribution over different environments. 
Besides, some research works explore RSSI sequential correlation to detect proximity.
Research work in~\cite{ProcedurallyGeneratedEnvironments} models RSSI sequence as a Markov chain and denotes the proximity as its model's binary state.
The work  of~\cite{shankarProximitySensingModeling2020a} experiments with several deep learning models on RSSI time series, trying to find a high dimensional classification boundary.
However, since multipath fading is complex and unpredictable, the resultant RSSI fluctuations are rather random and noisy.
Such fluctuation adversely affects these models, and, as a result, they cannot learn a good classification boundary.
In comparison, \n{} does not assume any noise distribution or sequential correlation. 
Since the multipath effect leads to a spread in RSSI distribution,
it represents the RSSI time series as a histogram and takes advantage of such signal fluctuation to extract environmental features.

Since different carriage states could render RSSI completely different even under the same environment and distance, 
some works further account for carriage state on BLE proximity detection.
The approach in~\cite{hatkeUsingBluetoothLow2020} categorizes carriage state as a known set of several on-body positions (such as being handheld or placed in a pocket).
They adjust their RSSI-based proximity threshold by carriage state while detecting proximity. 
However, this scheme relies heavily on manual labeling and works only for pre-defined states, making it hard to deploy or generalize in practice.
To efficiently reflect carriage state, the work in~\cite{shankarProximitySensingModeling2020a} introduces IMU and leverages deep learning to model the impact of carriage states on RSSI values.
To improve model generalizability, they crowdsource data from volunteers.
However, their models cannot generalize well since they have not considered the bias in the IMU training data.
As a comparison, \n{} uses IMU to reflect carriage state and addresses the bias issue by applying carriage state regularization. 
By accounting for the correlation between IMU readings and proximity states in training data, the regularization greatly reduces the IMU overfitting problem.

\section{Signal Encoding in \n{}}
\label{sec:system_feature}
We present in this section how \n{} processes RSSI and IMU signals so as to   encode the system inputs to feature vectors. In Section~\ref{subsec:histogramization}, we discuss histogramization for BLE RSSI. In Section~\ref{subsec:cfencode}, we introduce carriage feature encoding based on IMU readings. 

\subsection{RSSI Histogramization}
\label{subsec:histogramization}

\begin{figure} 
\centering
  \subfloat[Without pedestrian. \label{fig:sec3_hist1}]{%
        \includegraphics[width=0.49\linewidth]{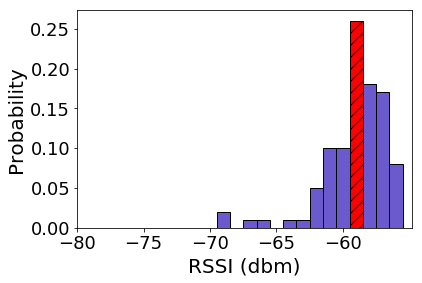}}
  \hfill
  \subfloat[With pedestrians.\label{fig:sec3_hist2}]{%
        \includegraphics[width=0.495\linewidth]{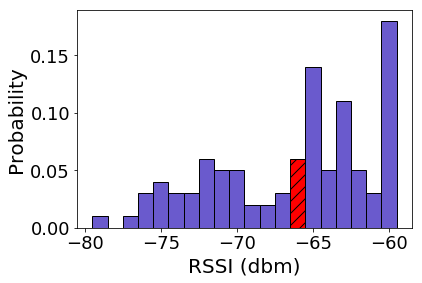}}
  \caption{BLE RSSI histogram under environments w/o pedestrians roaming around. The shaded red bin represents the median RSSI values.}
  \label{fig:sec3_hist} 
\end{figure}

Although an individual RSSI measurement is noisy, the distribution of RSSIs over time tends to be informative in representing environmental features.
Based on this observation, 
we use histogram to represent the distribution of RSSI fluctuation so as to capture RSSI distortion as well as the multipath in the environment. 
In the histogram, we present RSSI over a time period into various buckets, with each bucket denoting the appearance frequency of a certain RSSI range. 
Specifically, histogramization converts a series of RSSIs into a vector
\begin{equation}
    R=\left[r_1, r_2,...,r_n \right],
\end{equation}
where $r_i$ denotes the normalized number of RSSI (over a sliding window) that belongs to the $i^{th}$ bucket. 
We set each bucket to cover the same range length (bin size)
\begin{equation}
    \delta=\frac{\phi_{max}-\phi_{min}}{n},
\label{eq:bucket_range}
\end{equation}
where $\phi_{max}$ and $\phi_{min}$ represent RSSI extrema (maximum and minimum) that we consider, where, 
in our case, they are separately set to be $0\ dBm$ and $-100\ dBm$.

Figure~\ref{fig:sec3_hist} shows an example that illustrates the capability of RSSI histogram in representing RSSI distortions and multipath environment.
In the example, we collect RSSI in a corridor where the signal receiver is 3 meters from a transmitter, and the collection lasts for 10 seconds~(around 90 RSSI values are received).
Figure~\ref{fig:sec3_hist1} is collected at night when no pedestrian passes; and Figure~\ref{fig:sec3_hist2} is collected on a busy afternoon with people walking around.
Due to the multipath environments causing RSSI distortions to different extents, RSSI median value~(denoted as the red shaded bin) shifts between two figures without distance change, which violates the assumption that a lower RSSI means a larger distance and fails its derivative methods.
Nonetheless, RSSI histogramization represents the RSSI time series as a signal fluctuation distribution that could account for such influences from multipath environments.
For one thing, from static to dynamic environments, as the more complex multipath effect~(as in the Figure~\ref{fig:sec3_hist2} case) renders a larger RSSI distortion (where RSSIs are sequentially noisy), it also causes a larger signal fluctuation that leads to a spread in RSSI distribution. For another, the fluctuation enlarges the inconsistency between histogram buckets so that its shape tends to be disordered. 
Besides the implication of RSSI distortion and the multipath, the histogram also contains distance information because the buckets are gathered by the RSSI range. 

In our findings, this observation is also applicable to many non-line of sight and even cross-site scenarios.
This is because a complex environment is likewise to cause a complex fading effect, rendering a severe RSSI distortion with a large signal fluctuation.
This shows that the RSSI histogram is a powerful feature that is capable of representing multipath environments.



\subsection{Carriage Feature Encoding}
\label{subsec:cfencode}
We extract several features from IMU measurements to reflect carriage states.
Generally, IMU includes three categories of signals: gravity, linear acceleration, and angular velocity. 
Gravity can be naturally used to infer device attitude.
Linear acceleration and angular velocity, on the other hand, reflect the movement of a device.
We argue that the device movement is also an important factor in reflecting carriage states.
This is because these signals can further differentiate carriage states when devices have similar attitudes. 
For instance, the angular velocity along phone azimuth is much larger than its pitch when its user walks around with a phone in the front trouser pocket, while this situation reverses if the phone is in back pocket; in reality, such two carriage states could have different impacts on RSSI due to their different on-body positions.

We extract carriage features from a period of IMU data (of a few seconds).
Formally, we denote the feature vector by 
\begin{equation}
    C=\left[c_1, c_2, ..., c_m\right].
\end{equation}
For gravity, we directly average the values along its three dimensions -- the gravity projections of a device's 3D coordinates -- to reflect device attitude over time. 
For linear acceleration and angular velocity, we encode them by extracting some statistical features: energy, variance, skewness, kurtosis, and entropy.
Since feature extraction on IMU data is not the focus of this work, we empirically employ these features to reflect carriage states and summarize them in Table~\ref{tb:cf}.



\renewcommand\arraystretch{1.8}
\begin{table}
\center
\caption{Carriage feature elements extracted from IMU data.}
    \begin{threeparttable}[t]
    \begin{tabular}{|c|c|c|} 
        \hline 
            {\bf Feature} & {\bf Formula} & {\bf Source}${}^*$\\
        \hline
        \hline  
            Mean & $\frac{1}{N}\sum_{i=1}^N x_i$ & G\\
        \hline
            Energy & $\frac{1}{N} \sum_{i=1}^N {x_i}^2$ & L, A \\
        \hline 
            Variance & $\frac{1}{N} \sum_{i=1}^N (x_i-\bar{x})^2$ & L, A\\
        \hline
            Skewness & $E[(\frac{x-\bar{x}}{\sigma})^3]$ & L, A \\
        \hline
            Kurtosis & $E[(\frac{x-\bar{x}}{\sigma})^4]$ & L, A \\
        \hline
            \multirow{2}*{Entropy} & $E[I(x)], $ & \multirow{2}*{L, A}\\
                  ~  & where $I(x)=-\ln\left[Pr\left(\frac{x-\min(x)}{\max(x)-\min(x)}\right)\right]$& \\
        \hline
    \end{tabular}
    \label{tb:cf}
    \footnotesize{* G: gravity, L:linear acceleration, A: angular velocity}
    \end{threeparttable}
\end{table}

\section{Proximity Classification in \n{}}
\label{sec:system_classification}
We discuss
in this section how \n{} detects proximity given 
the feature vectors of RSSIs and carriage states. 
In Section~\ref{subsec:classifier}, we present the design of the DNN-based proximity classifier. In Section~\ref{subsec:cfr}, we discuss carriage state regularization for training the classifier.

\subsection{DNN-based Binary Proximity Classifier}
\label{subsec:classifier}
As mentioned, RSSI histogram $R$ provides environment and distance features, while carriage feature vector $C$ reflects carriage state. 
We need a classification model that jointly considers these inputs to estimate 
\begin{equation}
\label{eq:class_model}
    y=Pr\left(Y=1|R,C\right),
\end{equation}
where $y$ is the classifier output regarding the proximity state~$Y$.
As mentioned, proximity state is either "proximity" event (1) or "no proximity" event (0), i.e., 
\begin{equation}
\label{equa:cal_prox}
    Y =
    \begin{dcases}
        1, & \text{if } D \leq \tau \,;\\
        0, & \text{otherwise} \,, \\
    \end{dcases}
\end{equation}
where $\tau$ is the pre-defined proximity threshold and $D$ is the physical distance.

Unfortunately, it is intractable to handcraft such a model with those high-dimensional inputs. 
Therefore, we leverage deep learning to extract features from those inputs and correlate them with the proximity state.
Specifically, the RSSI histogram and the carriage features, which are encoded over the same period, are concatenated as a feature vector.
We then use a deep neural network (DNN)-based proximity classifier to determine proximity states from the feature vectors.

\begin{figure}[t]
\centering
\includegraphics[width=0.4\textwidth]{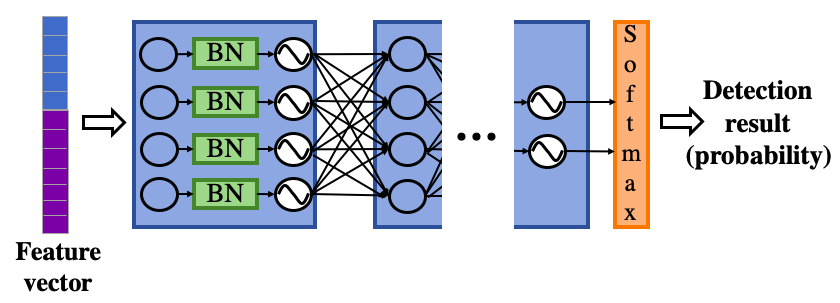}
\caption{
The network structure of the proximity classifier in \n{}. 
}
\label{fig:prox_class}
\vspace{-0.1in}
\end{figure}

We illustrate the classifier structure in Figure~\ref{fig:prox_class}. 
It consists of an input layer (omitted in the figure), an output layer, and several hidden layers. 
Each hidden layer is comprised of network neurons, a normalization method, and an activation function.
We employ batch normalization (BN)~\cite{ioffeBatchNormalizationAccelerating2015} as the normalization method and Mish~\cite{misraMishSelfRegularized2020} as the activation function.
As a convention, Softmax serves as the classifier's output layer.
As for the model training, we choose cross-entropy as the loss function and train the network with the Adam optimizer. 
We have also tried other common deep learning model tricks, such as using ReLU as the activation function; they show a similar performance in this task.
Readers interested in such tricks may refer to~\cite{apicellaSurveyModernTrainable2021,Goodfellow-et-al-2016, chenJointDemosaickingDenoising2021} and the references therein.

\subsection{Carriage State Regularization}
\label{subsec:cfr}

\begin{figure}[t]
\centering
\includegraphics[width=0.32\textwidth]{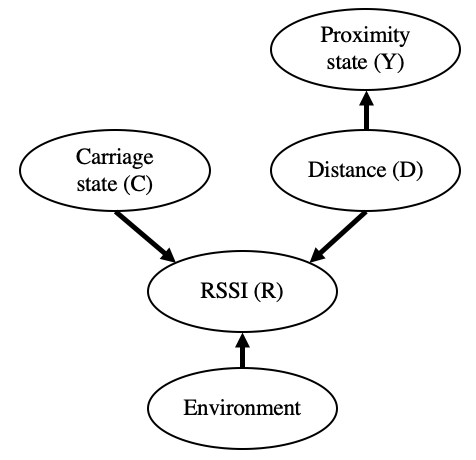}
\caption{
Graphical model of BLE proximity detection.
Each node denotes a factor or event in proximity detection.
A directed edge denotes causality between its connected nodes.
}
\label{fig:cfr_causality}
\vspace{-0.1in}
\end{figure}

We illustrate in Figure~\ref{fig:cfr_causality} the problem of BLE proximity detection by posing it as a graph model.
Each node represents a factor or an event in the BLE proximity detection system.
In the figure, the considerations -- carriage state, distance, and multipath environment -- are three major factors that influence RSSI; proximity state is the comparison result between proximity threshold (omitted in the graph for simplicity) and physical distance.
We denote RSSI with all of its representation as $R$ and IMU reading with its derived carriage state features as $C$.
Take \n{} as an example, we denote $R$ as an RSSI histogram and $C$ as carriage feature vector.

In the figure, a directed edge describes causality between nodes.
Specifically, a directed edge pointing A to B indicates that a variation in A could lead to a status change in B, being other factors unchanged.
For example, an edge pointing from distance to RSSI means: without other factors changing, a larger distance would lead to a smaller RSSI value.
Note that, the association absence between any two nodes assumes their prior independence;
for instance, carriage state and distance are independent being RSSI unknown, but they are correlated when the RSSI is given~(according to d-separation~\cite{10.1214/09-SS057}).

Essentially, the BLE proximity detection task is to model those causalities beforehand, so that we can use the established model to estimate the proximity state from those observable inputs. 
For regression works that only consider distance and RSSI, they model $Pr(R|D)$ because distance causes RSSI change.
In practice, they estimate the distance from the observable RSSI  
\begin{equation}
    Pr(D|R)\propto Pr(R|D),
\end{equation}
assuming $Pr(D)$ is uniformly distributed.
Then, proximity state can be estimated the same way as in Equation~\ref{equa:cal_prox}, with the goal of modeling the causality from distance to proximity state in Figure~\ref{fig:cfr_causality}.
Apparently, these works suffer inaccuracy in practice due to the absence of carriage state and environment as well as their related edges.
Thus, later works further consider environment and carriage state, aiming to model $Pr(R|C,Y)$ in the training phase.
Since $Pr(R|C,Y)$ is intractable, they leverage deep learning to directly learn $Pr(Y|R,C)$ from data, assuming that
\begin{equation}
\label{eq:approx}
    Pr(Y|C,R)\propto Pr(R|Y,C).
\end{equation}

However, Equation~\ref{eq:approx} is valid only if proximity state $Y$ and carriage state $C$ are independent because
\begin{equation}
    Pr(Y|C,R)=\frac{Pr(Y,C,R)}{Pr(C,R)}\propto Pr(R|Y,C)Pr(Y|C). 
\end{equation}
For using non-parametric models as deep learning ones, we expect this independence in training data to guarantee the established model to make unbiased estimations in practice. 
In other words, instead of determining proximity from carriage state, we want the model to detect proximity based on RSSI representations while considering carriage state impact on RSSI.
Unfortunately, it is usually tricky to guarantee such unbiased training data in quantity -- for one thing, IMU is highly sensitive to user activities, which varies in different scenarios. For another, acquiring such data is labor-intensive, systems from the literature (such as the work in~\cite{tc4tlChallenge2020}) crowdsource data from volunteers for better adaptability in different environments; this uncontrolled data acquisition process makes it harder to guarantee an unbiased training dataset. 
As a result, the built model $Pr(Y|C,R)$ suffers bias from $Pr(R|Y,C)$, and its performance is highly manipulated by training data.
We refer to this issue as an IMU overfitting problem because the learned model falsely correlates IMU with proximity state when this discrepancy from independence appears in the training data.
Apparently, this issue hinders the models from generalizing well.

To tackle this IMU overfitting problem, we propose carriage state regularization to supervise the training process of the DNN-based proximity classifier.
Intuitively, carriage state regularization aims to cut off the correlation between the carriage feature and proximity state from the training data, so as to force the proximity classifier to learn the carriage state's impact on RSSI. 

Carriage state regularization leverages importance sampling to account for the distribution discrepancy of the carriage feature with different proximity states. 
We divide the training dataset $D$ into two groups $S= S_{Y=0} \cup S_{Y=1}$, 
according to their proximity state $Y$.
The discrepancy is caused by the fact that these two groups are drawn from different distributions so that
    $Pr(C|Y=1)\ne Pr(C|Y=0)$.
Such a discrepancy can be reduced by reweighting one group to match the other.
Thus, with the goal to tackle the resultant overfitting problem, we reweight the training loss to reduce the discrepancy by
\begin{equation}
    L'(y_i,Y_i)=\frac{Pr(C_i|Y=1)}{Pr(C_i|Y=0)}L(y_i,Y_i)=w_iL(y_i,Y_i),
\end{equation}
where $L(\cdot)$ calculates training loss from detection result $y_i$ and proximity state $Y_i$, and $w_i$ represents the sampling weight.

Let $\boldsymbol{w}^{+}$ be the set of weights corresponding to the positive samples, i.e., 
$\boldsymbol{w}^{+} = \left\{ w_i | s_i \in S_{Y=1} \right\}$.
Similarly, the weight set for negative ones are $\boldsymbol{w}^{-} = \left\{ w_i | s_i \in S_{Y=0} \right\}$.
In this work, we aim to reweight $S_{Y=1}$ to match $S_{Y=0}$. 
We thus set all $w_i \in \boldsymbol{w}^{-}$ to be 1 and calculate $\boldsymbol{w}^{+}$.
We use the kernel method to bridge the distribution discrepancy in the feature space~(kernel mean matching~\cite{huangCorrectingSampleSelection2006b}), that is 
\begin{align}
&\min_{\boldsymbol{w^+}} \quad \frac{1}{2}(\boldsymbol{w^+})^TK\boldsymbol{w^+}
  -\kappa^T\boldsymbol{w^+} \\
& \begin{array}{r@{\quad}l@{}l@{\quad}l}
s.t. & \sum_{i=1}^{||\boldsymbol{w^+}||} w^+_i = ||\boldsymbol{w^+}|| \\
& 0 \leq w^+_i\leq w_{max},\ i=1,2,3\ldots, ||\boldsymbol{w^+}||  
\end{array},
\end{align}
where we set $w_{max}$ as 10. The kernel computations are
\begin{align}
    K_{ij}&:=k(C^+_i, C^+_j), \\
    \kappa_i&:=\frac{||\boldsymbol{w^+}||}{||\boldsymbol{w^-}||}\sum_{j=1}^{||w^-||}k(C^+_i,C^-_j),
\end{align}
where $k(\cdot,\cdot)$ is the radial basis function~(RBF) kernel (the kernel width $\gamma=1.0$);
we solve this quadratic program problem by the interior point method~\cite{Boyd2006CO}.

Note that carriage state regularization is general for IMU-assisted BLE proximity detection. Although we explain this regularization on the carriage feature, it is also applicable to other features that reflect carriage states.

\section{\n{}-lite: Achieving Memory Efficiency}
\label{sec:PRID-Lite}
\begin{figure}[t]
\centering
\includegraphics[width=0.48\textwidth]{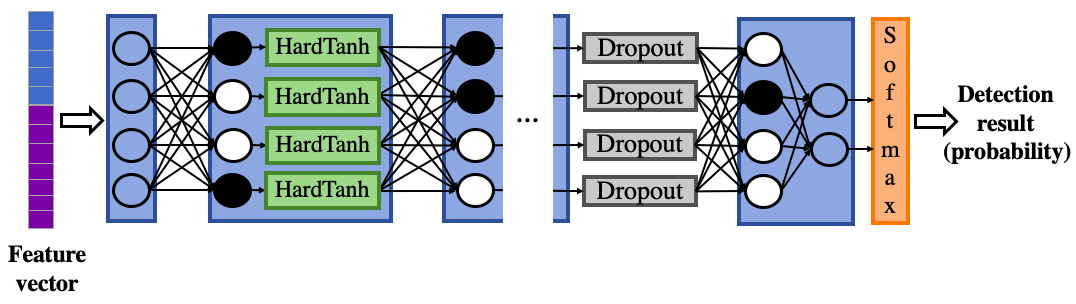}
\caption{The network structure of proximity classifier in \n{}-lite.}
\label{fig:bnn_pipeline}
\end{figure}

In \n{}, most of the storage burden comes from the DNN-based proximity classifier. 
Normally, we use the larger neural network (with more neurons) for better generalization.
However, a larger neural network not only requires more memory but also increased energy consumption that some resource-constraint devices cannot afford.
Thus, the constrained resource limits us from using a large neural network, which hinders the proximity classifier from being accurate.

Since a neural network's nature is to encode training data into its neuron connections, the neural network needs to be large enough to perform well; thus, we apply binarized neurons to build a larger network when the model size is constrained.
Instead of using a floating-point number, a binarized neuron weight is either 1 or -1 (represented by a bit). 
This allows us to replace a full-precision neuron to be 32 binarized neurons, which enlarges the model structure in terms of neuron quantity.
In addition, this binarization replaces the floating-point multiplication between neurons by bit-wise operation, which is faster and lighter for IoT devices to process.
Although this binarization causes information loss between neurons -- which inevitably renders inaccuracy of binarized neural network in comparison to its full-precision counterpart with enough neuron quantity -- it is likely to gain a better performance by compromising the neuron precision for a larger network structure when device resources limit the scaling up of the model.

We illustrate the binarized proximity classifier structure in Figure~\ref{fig:bnn_pipeline}.
Compared with the proximity classifier from \n{}, we binarize parts of its neurons (colored in white and black) and apply the activation function $\mbox{HardTanh}$ to the hidden layers.
We maintain the neuron precision of the input and last hidden layer because they are important in conveying information; but this would not cause severe memory burdens since these two layers are usually much smaller than the other layers.
For general regularization, we add one dropout layer between the last two hidden layers.
We train the network by using the same tricks as in the proximity classifier in \n{}, except that we employ ``straight-through estimator''~\cite{courbariauxBinarizedNeuralNetworks2016} to tackle the binarized neuron training problem.

\section{Illustrative Experimental Results}
\label{sec:exp}
In this section, we show the illustrative experimental results of \n{} and \n{}-lite. We first discuss the experimental settings
in Section~\ref{subsec:exp_setting}, followed by the
illustrative results 
in Section~\ref{subsec:illus_result}.

\subsection{Experimental Settings}
\label{subsec:exp_setting}

We have implemented \n{} and \n{}-lite and conducted extensive experiments to validate their performance.
We run \n{} on an Android smartphone, which continuously scans any surrounding BLE signals and logs the IMU data. 
For \n{}-lite, we deploy our implementation on multiple resource-constrained IoT devices, including an Android smartwatch, a single-board computer Raspberry Pi Zero (Pi Zero), and a microcontroller unit ESP-32. 
Detailed specifications of the IoT devices are listed in Table~\ref{tb:res_hardware}. 

In our implementation of the DNN-based binary proximity classifier, we employ three fully connected layers with 700, 900, and 700 hidden nodes. Unless specified, it is for both \n{} and \n{}-lite.
\begin{figure}
\centering
\includegraphics[width=0.32\textwidth]{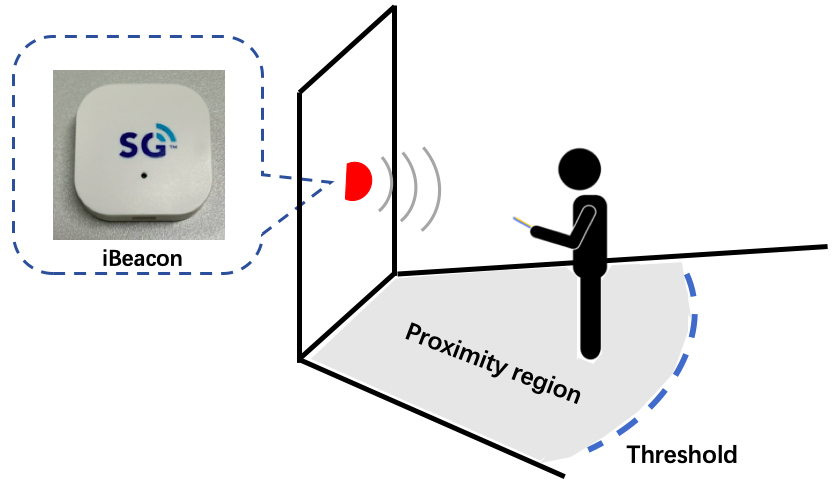}
\caption{Illustration of data collection. The dotted line stands for proximity threshold. The user either moves within the proximity region or outside the proximity threshold.}
\label{fig:exp_illu}
\end{figure}

We verify the system performance through an experiment on fields and an evaluation on a public crowdsourced dataset~\cite{tc4tlChallenge2020}. 
In our experiment, we set up a commercial iBeacon as the BLE transmitter. 
The iBeacon is attached to a wall 1.2 meters above the ground (as illustrated in Figure~\ref{fig:exp_illu}). It advertises 10 BLE handshaking messages per second with reference TX power (RSSI measured at 1 meter) of -59dBm. 
We invite multiple users for data collection. 
They carry smartphones either by hands or in pockets (decided by the users) and can move within 12m of the transmitter. 
We collect data in three venues: a crowded indoor junction (dynamic environment), an indoor open space with a few pedestrians (semi-dynamic environment), and a quiet outdoor region (static environment). 
In total, we collect RSSI readings for around 5 hours for model training and another one hour for performance testing.
The detailed collecting hour of training data is shown in Table~\ref{tb:cs_data}.
To fairly reflect reality, the carriage state and proximity state labels are carefully decoupled in the test data.
In addition, we repeat the experiments for the proximity thresholds of 6m and 2m, corresponding to different proximity requirements in typical PBS applications (such as proximity marketing) and contact tracing scenarios, respectively. 

We also validate our scheme on a public dataset from TC4TL challenge~\cite{tc4tlChallenge2020}, a BLE proximity detection competition for automated contact tracing. 
The TC4TL dataset contains more than 25,000 crowdsourced RSSI sequences with each sequence ranging from 10 to 150 seconds.
It covers several common environments and carriage states for contact tracing and restricts users to move within 4.5m from transmitters.
We use the 2m proximity threshold in the evaluation since it is a common safe distance for contact tracing.

For a better evaluation, we separately train and test all the methods (including comparison schemes and our method) on the two datasets. We chunk the test data into time periods of 5s as well as the sliding window length. The evaluation metrics of this experiment are shown as follows:
\begin{itemize}
    \item \emph{Accuracy:} Accuracy is a common and intuitive metric for evaluating classifier quality. 
    It denotes the proximity state as $Y$ and detection result as $Det$.
    The system precision and recall is calculated as
    \begin{equation}
       \begin{aligned}
       Precision=\frac{||\{Y=1\} \cap \{Det=1\}||}{||\{Det=1\}||}, \\
       Recall=\frac{||\{Y=1\} \cap \{Det=1\}||}{||\{Y=1\}||}.
       \end{aligned}
    \end{equation}
    F-score is their harmonic mean, which is computed by
    \begin{equation}
        F_1=\frac{2 \times Precision \times Recall}{Precision + Recall}.
    \end{equation} 
    
    \item \emph{False detection rate:} 
    Accuracy mainly focuses on system discernment on positive events, while recognizing proximity events and reducing false alarms are both essential in many PBS applications (such as contact tracing); thus,  
    we further use a false detection rate to compare system performance. 
    We follow the metric in~\cite{shankarProximitySensingModeling2020a} and employ nDCF to measure the false detection rate.
    We evaluate the detection error as the probability of missing contact (proximity) event $E_{miss}$ and false alarm $E_{fa}$, which are computed by 
    \begin{equation}
        \begin{aligned}
            E_{miss}&=&\frac{||\{Y=1\}\cap \{Det=0\}||}{||\{Y=1\}||},\\
            E_{fa}&=&\frac{||\{Y=0\} \cap \{Det=1\}||}{||\{Y=0\}||}.
        \end{aligned}
    \end{equation}
    The evaluation nDCF is their normalized decision cost function
    \begin{equation}
        nDCF=\frac{w_{miss}E_{miss}+w_{fa}E_{fa}}{\min(w_{miss},w_{fa})},
    \end{equation}
    where we use weights $w_{miss}=1$ and $w_{fa}=1$ in this experiment.
\end{itemize}

We compare our scheme with the following state-of-the-art schemes:
\begin{itemize}
\item \emph{Temporal 1-D convolutional network (Conv1D)}~\cite{shankarProximitySensingModeling2020a}:
Conv1D is a deep learning-based binary regressor. 
It uses raw RSSI and IMU data over a time period as input. 
In the experiments, we use one convolutional (with kernel size of $1\times 5$) and three fully-connected layers.
Its neuron quantity is set to be similar to that of \n{}.

\item \emph{ProxiTrak}~\cite{chandelProxiTrakRobustSolution2020}:
ProxiTrak employs a random forest to detect proximity events. 
It extracts mean, minimum, maximum, and standard deviation from both the RSSI sliding window and inter-packet duration (IPD). 
It further conducts majority voting on the detection results to mitigate random noise. 
In our experiments, we follow the original paper to set the sliding window's length to be 2s, and then aggregate results in 5s.

\item \emph{Log-distance path loss model (LDPL)}~\cite{alqathradyImprovingBLEDistance2017}: 
LDPL is a classic distance estimation technique and wildly used in commercial regression-based proximity detection systems. 
The distance is estimated by
\begin{equation}
    distance=10^{\frac{TX-RSSI}{10n}},
\end{equation} 
where $n$ is a fading parameter learned from training data and TX denotes the reference RSSI measured at 1m.   
It detects proximity events by comparing the resultant distance with the proximity threshold.
To alleviate signal fluctuation, we calculate RSSI by the mean value over a sliding window of 5s. 
\end{itemize}


\subsection{Illustrative Results}
\label{subsec:illus_result}

\begin{figure*}[htbp]
\centering
\begin{minipage}[t]{0.3\textwidth}
\centering
\includegraphics[width=1.05\linewidth]{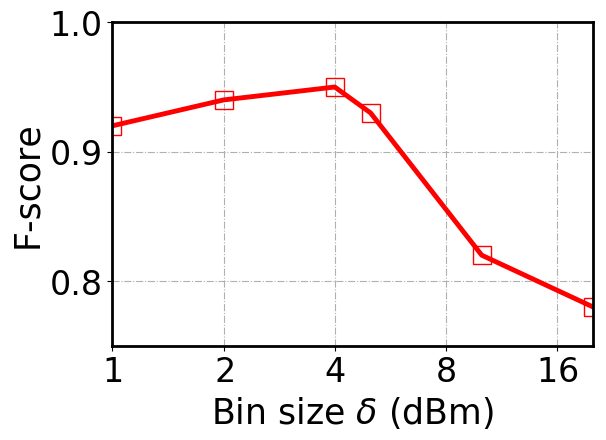}
\caption{RSSI histogram bin size setting.}
\label{fig:bin_size}
\end{minipage}
\hspace{0.15in}
\begin{minipage}[t]{0.3\textwidth}
\centering
\includegraphics[width=\linewidth]{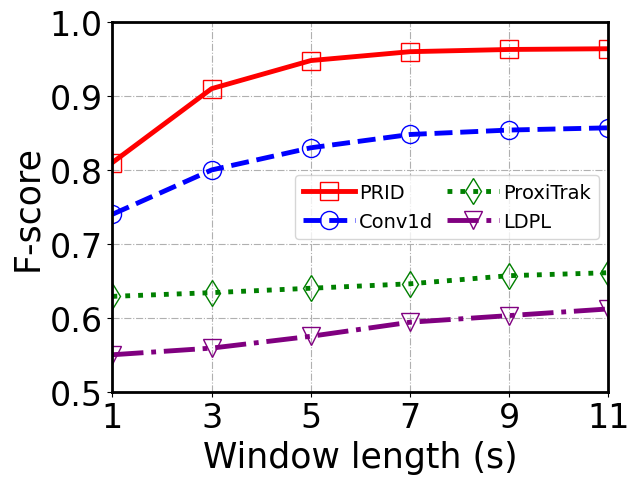}
\caption{F-score against sliding window length.}
\label{fig:win_length}
\end{minipage}
\hspace{0.15in}
\begin{minipage}[t]{0.28\textwidth}
\centering
\includegraphics[width=1.05\linewidth]{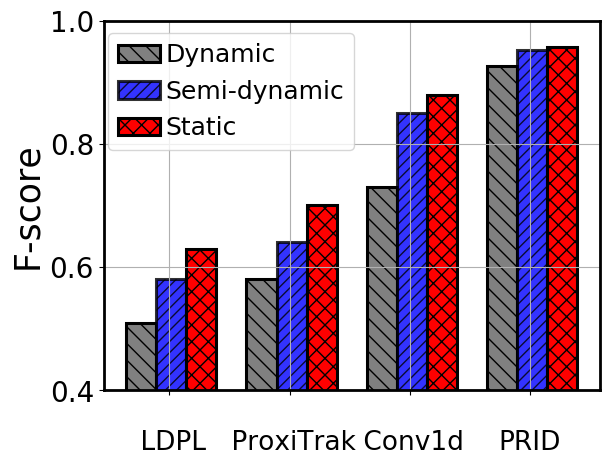}
\caption{F-score under different sites.}
\label{fig:site_noise}
\end{minipage}
\end{figure*}

\begin{figure*}
\centering
    \begin{minipage}[t]{0.3\textwidth}
    \centering
    \includegraphics[width=\linewidth]{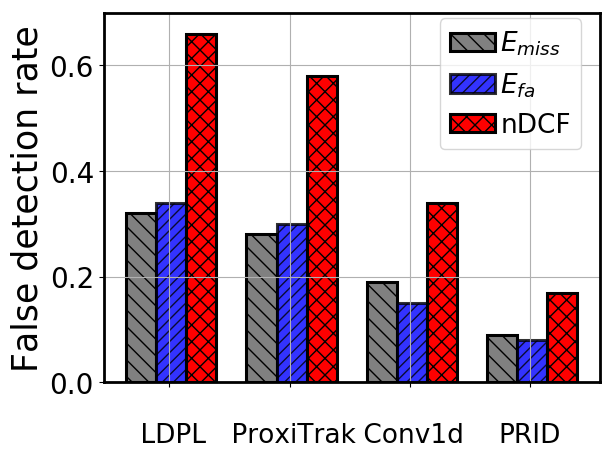}
    \caption{Classification performance for contact tracing application (proximity threshold is 2m).}
    \label{fig:thres2_exp}
    \end{minipage}
    \hspace{0.1in}
    \begin{minipage}[t]{0.3\textwidth}
        \centering
        \includegraphics[width=\textwidth]{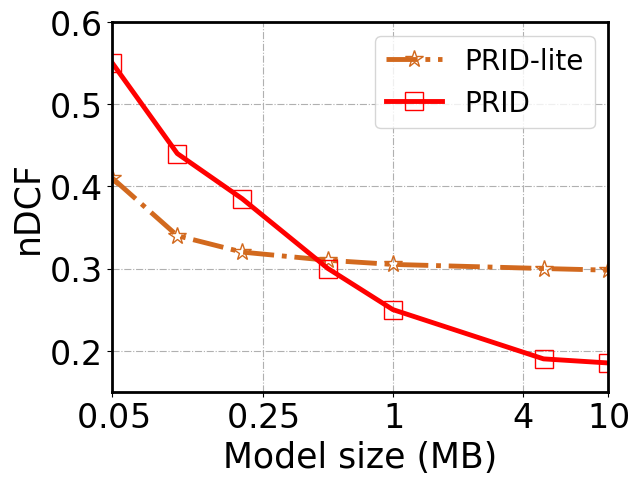}
        \caption{nDCF against different model size of \n{} and \n{}-lite. }
        \label{fig:lite_modelsize}
    \end{minipage}
    \hspace{0.1in}
    \begin{minipage}[t]{0.3\textwidth}
        \centering
        \includegraphics[width=.95\textwidth]{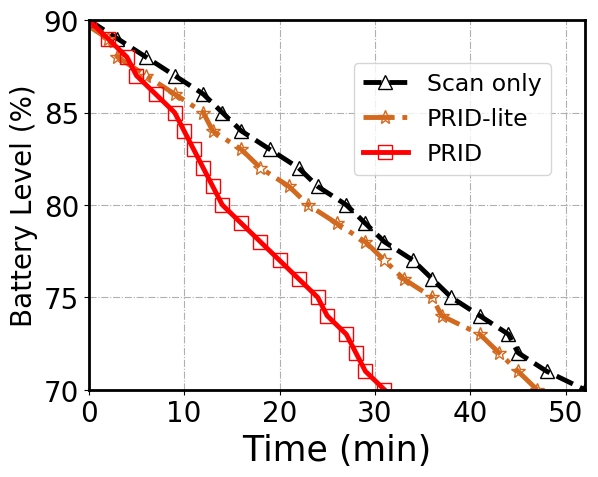}
        \caption{Power consumption over time on IoT smartwatch.}
        \label{fig:lite_power}
    \end{minipage}
\end{figure*}

\renewcommand\arraystretch{1.5}
\begin{table}
\center
\caption{Scheme precision/recall under different sites.}
\begin{minipage}[t]{\columnwidth}
    \centering
    \resizebox{\columnwidth}{!}{
    \begin{tabular}{|c|c|p{3.5mm}p{3.5mm}|ccc|}
        \hline
        \multicolumn{4}{|c|}{\bf Collection} & \multicolumn{3}{c|}{\bf Scheme precision/recall}\\
        \hline
        \hline
         Environment& Carriage & Y=0 & Y=1 & LDPL & Conv1d & \n{} \\
         \hline
         \multirow{2}*{Dynamic} & Pocket & 0.48 & 0.30 & 0.64/0.62  & 0.88/0.76 & 0.95/0.92 \\
         \cline{2-7}
         ~ & Hand & 0.32 & 0.5 & 0.63/0.64 & 0.78/0.88 & 0.93/0.95 \\
         \hline
         \multirow{2}*{Semi-dynamic} & Pocket & 0.36 & 0.45 & 0.65/0.66 & 0.83/0.92 & 0.94/0.95\\
         \cline{2-7}
         ~ & Hand & 0.48 & 0.40 & 0.68/0.70 & 0.91/0.82 & 0.95/0.95 \\
         \hline
         \multirow{2}*{Static} & Pocket & 0.42 & 0.38 & 0.70/0.72 & 0.91/0.85 & 0.96/0.95 \\
         \cline{2-7}
         ~ & Hand & 0.38 & 0.42 & 0.72/0.71 & 0.86/0.91 & 0.96/0.97 \\
         \hline
        \end{tabular}
    }
    \end{minipage}
    \label{tb:cs_data}
\end{table}

We first study the impact of histogram bin size $\delta$ (as in Equation~\ref{eq:bucket_range}) on the F-score.
Figure~\ref{fig:bin_size} shows that the F-score increases with the bin size when $\delta \le 4$; then, the F-score drops when bin size increases.
This is because that histogram with a larger bin is less sparse, providing stable features to the proximity classifier to capture; 
but a histogram with a too large of a bin size is insensitive to RSSI fluctuation distribution change, so it cannot represent the multipath environment well.
Therefore, we follow the figure trend and adopt $\delta=4$ in the following experiments.

Figure~\ref{fig:win_length} compares the F-score of different schemes under varied window lengths (in terms of time) in our on-field experiment. 
We train and test data from the three sites together.
In the figure, \n{} outperforms the other schemes by more than 0.13 in the F-score (or reduces in over 60\% of the false detection cases). 
We can see that \n{} and Conv1d gain significant improvements when the window length increases from 1s to 5s, then after that, they level out.
ProxiTrak and LDPL, on the other hand, do not benefit much from temporal information.
This is because the RSSI sliding window contains environmental features, and the DNN-based methods are able to extract them.
Thus, to balance the system accuracy and responsiveness, we set the window length to be 5s in both \n{} and \n{}-lite.

With the 5s window, we further show the system performance upon the three sites in Figure~\ref{fig:site_noise}.
As mentioned, the dynamic environment is crowded with pedestrians, the semi-dynamic environment is relatively less crowded, and the static site has quiet surroundings;
Therefore, these sites possess environmental dynamics and hence some multipath effects to different extents.
As a result, all the schemes suffer accuracy loss from static to dynamic environments because of RSSI fluctuation.
Nonetheless, across different sites, the negligible accuracy changes of \n{} shows its robustness to RSSI fluctuation since it captures the multipath environment by RSSI histogramization.

To study the system's robustness against IMU data bias, we separately conduct experiments on the three sites.
We show scheme precision and recall in Table~\ref{tb:cs_data}.
Since ProxiTrak has a similar trend to LDPL, we only show \n{}, Conv1d, and LDPL to avoid redundancy.
Due to the multiple collectors and different surroundings, carriage state and proximity state are not independent in each site. 
From the collection hour shown in the table, the carriage state strongly correlates with the proximity state in the dynamic environment site. This correlation becomes less in the semi-dynamic site and close to being independent in the static site. 
All the schemes perform better than those training on the three sites together,
because training on one specific site makes the models more customized.
From the table, LDPL shows inaccuracy since it cannot adapt to the multipath environments and different carriage states.
Although Conv1d gets commendable results by considering the carriage state, its accuracy (in terms of precision and recall) is manipulated by IMU data bias.
\n{} achieves high accuracy and robustness in all sites since it employs the carriage state regularization to tackle IMU overfitting; even in a static site where the carriage and proximity states are close to being decoupled, \n{} is still superior to Conv1d. 
This is because IMU readings reflect device poses, its bias could still exist even though the carriage state is nearly balanced in the training data.
Overall, this table shows that \n{} is not only robust to IMU data bias but highly reliable under various sites.

Contact tracing is an essential application of proximity detection. 
Unlike classic PBS systems, contact tracing requires a smaller proximity threshold (2m), rendering it very challenging. 
To validate system performance in this application, as well as verify the generality (in terms of proximity threshold) of \n{}, we conduct experiments under the proximity threshold of 2m. 
For fair evaluation, we consider the cases collected within a 4-meter distance for training and testing. 
Figure~\ref{fig:thres2_exp} demonstrates the nDCF of different schemes. 
All the schemes suffer slight performance declines when the proximity threshold is 2m instead of 6m.
This is because the user moving range shrinks from 12m to 4m, introducing severe signal ambiguity -- 
the smaller proximity threshold enlarges the influence of environmental dynamics and carriage states. 
Nonetheless, \n{} reduces the false detection by more than 50\% (in terms of nDCF) compared with other schemes.

We evaluate the system performance in terms of nDCF on the TC4TL challenge dataset with our comparison schemes: LDPL (0.82), ProxiTrak (0.73), Conv1d (0.58), and \n{} (0.49). 
In this dataset, since several transmission periods share one ground truth, we apply a majority voting to assemble results from several periods.
This dataset is challenging because it crowdsources data from large numbers of users with diverse environments, with its training and test data not necessarily from the same sets of environments or carriage states.
It requires not only good adaptability upon diverse environments and carriage states but also generalization ability to new scenarios.
Nevertheless, the results show that our scheme achieves the best results among the comparison schemes by reducing more than 15\% false detection cases compared with the existing arts.
This is because \n{} extracts better features from inputs and considers the IMU overfitting problem.
This result also verifies the better practicability and reliability of \n{} in reality.

\renewcommand\arraystretch{1.5}
\begin{table}
\center
\caption{IoT hardware specification and computational cost of \n{}-Lite.}
    \centering
    \begin{tabular}{|c|c|c|c|c|}
        \hline
         \multirow{2}*{Hardware} & \multicolumn{2}{c|}{Specification} & \multicolumn{2}{c|}{Computational cost} \\
         \cline{2-5}
          ~ & CPU Freq. & RAM & Time & Memory usage \\
        \hline
        \hline
        Smartwatch & 1.5GHz& 3GB& $<$1ms& 318KB\\
        \hline 
        Pi Zero & 1GHz& 512MB& 4ms& 314KB\\
        \hline
        ESP-32 & 240MHz& 512KB& 268ms& 309KB\\
        \hline
        \end{tabular}
        \label{tb:res_hardware}
    \label{tb:para}
\end{table}

To deploy \n{} on extremely resource-constrained devices, we study the system's performance (in terms of nDCF) with different proximity classifier model sizes of \n{} and \n{}-lite. It is conducted on our on-field experiment when the proximity threshold is 2m.
From Figure~\ref{fig:lite_modelsize}, \n{} nDCF first drops as the model is smaller than $\sim$5MB and then levels out after that. 
While nDCF of \n{}-lite levels out when model is larger than 0.3MB.
Note that, \n{}-lite outperforms \n{} when the model size is less than 0.5MB. 
This is because the neuron quantity of \n{} is so small that the model capacity and generalization ability are constrained.
In contrast, \n{}-lite sacrifices a percentage of neuron precision for neuron quantity, so it achieves a lower nDCF when the model size is small.
However, the binarized neuron causes information loss between network layers, making it hard for \n{}-lite to perform as accurately as \n{} does when the model size is large enough.
From the figure, we adopt the model size of 5MB and 0.3MB for \n{} and \n{}-lite in our experiment. 
It shows that \n{}-lite reduces 94\% memory cost at the expense of 7\% accuracy loss in F-score compared with \n{}.

We evaluate the energy consumption over time on \n{} and \n{}-lite, using an IoT smartwatch.
The watch is equipped with a lithium polymer built-in battery with a capacity of 800mAh. 
Except for BLE scanning and proximity detection, we kill all unnecessary processes. 
The smartwatch conducts one BLE scan per 10 seconds, and each scan lasts for 5 seconds. 
We record battery levels from 90\% to 70\% and show them in Figure~\ref{fig:lite_power}. 
From the figure, the BLE scan consumes 20\% battery energy within around 50 minutes. 
Scanning while running the \n{} takes 30 minutes to use up the same amount of battery power.
In comparison, \n{}-lite is more energy-efficient as it consumes negligible power except for BLE scanning. 
Overall, \n{}-lite extends battery life by 60\% compared to \n{}, making it efficient for energy-constrained IoT devices in our experiment. 

We finally show in Table~\ref{tb:res_hardware} the inference time and memory usage of \n{}-lite on different IoT devices. 
We run \n{}-lite 10,000 times on each device and the average time cost as inference time that affects system responsiveness.
From the table, even on ESP-32, the platform with the lowest computational power among the devices, \n{}-lite costs only 0.3s for each detection.
As for memory usage, \n{}-lite takes up $\sim$0.3MB memory on all three devices.

\section{Conclusion}
\label{sec:conclude}
BLE-based proximity detection plays important roles in many proximity-based services. 
It relies on the fact that a lower RSSI implies a longer distance, and vice versa.
However, RSSI can be markedly affected by multipath and device carriage states in reality.
Previous works in the area 
have not sufficiently considered RSSI fluctuation due to multipath, and how to address carriage states with IMU training data bias to achieve highly accurate  and robust proximity detection.
%
%
%

We propose \n{}, a novel IMU-assisted BLE proximity detection approach robust against RSSI fluctuation due to multipath and IMU data bias.
By representing RSSI fluctuation as a histogram,
\n{} extracts the features of the multipath environment to encode it as a feature vector.  It further encodes the IMU data into carriage feature. Employing a DNN-based binary proximity classifier, \n{} then detects proximity after concatenating the vectors. To address biased training data,
 \n{} applies carriage state regularization to the loss function to reduce IMU overfitting of the DNN-based classifier.
To make our scheme more deployable on highly resource-constrained IoT devices, we further propose \n{}-lite, a lightweight version of \n{} using a binarized neural network.
We have implemented \n{} and \n{}-lite in IoT devices. We conduct extensive experiments in several venues and a public dataset and compare them with the state-of-the-arts.
Our results show that \n{} achieves  substantial improvement as compared with the state of the arts (with more than 50\% reduction in false detection cases in our dataset). 
\n{}-lite reduces the model size from \n{} by 94\% (from 5MB to 0.3MB)
and extends battery life by more than 60\%, with only a minor cost in accuracy  (7\% reduction in F-score).


\bibliographystyle{IEEEtran}
\bibliography{main}

\begin{thebibliography}{10}
\providecommand{\url}[1]{#1}
\csname url@samestyle\endcsname
\providecommand{\newblock}{\relax}
\providecommand{\bibinfo}[2]{#2}
\providecommand{\BIBentrySTDinterwordspacing}{\spaceskip=0pt\relax}
\providecommand{\BIBentryALTinterwordstretchfactor}{4}
\providecommand{\BIBentryALTinterwordspacing}{\spaceskip=\fontdimen2\font plus
\BIBentryALTinterwordstretchfactor\fontdimen3\font minus
  \fontdimen4\font\relax}
\providecommand{\BIBforeignlanguage}[2]{{%
\expandafter\ifx\csname l@#1\endcsname\relax
\typeout{** WARNING: IEEEtran.bst: No hyphenation pattern has been}%
\typeout{** loaded for the language `#1'. Using the pattern for}%
\typeout{** the default language instead.}%
\else
\language=\csname l@#1\endcsname
\fi
#2}}
\providecommand{\BIBdecl}{\relax}
\BIBdecl

\bibitem{ngCOVID19YourSmartphone2020}
P.~C. Ng, P.~Spachos, and K.~Plataniotis, ``{{COVID}}-19 and {{Your
  Smartphone}}: {{BLE}}-based {{Smart Contact Tracing}},''
  \emph{arXiv:2005.13754 [cs]}, May 2020.

\bibitem{li_spatial-temporal_2021}
\BIBentryALTinterwordspacing
G.~Li, C.-C. Hung, M.~Liu, L.~Pan, W.-C. Peng, and S.-H.~G. Chan,
  ``\BIBforeignlanguage{en}{Spatial-{Temporal} {Similarity} for {Trajectories}
  with {Location} {Noise} and {Sporadic} {Sampling}},'' in
  \emph{\BIBforeignlanguage{en}{2021 {IEEE} 37th {International} {Conference}
  on {Data} {Engineering} ({ICDE})}}.\hskip 1em plus 0.5em minus 0.4em\relax
  Chania, Greece: IEEE, Apr. 2021, pp. 1224--1235. [Online]. Available:
  \url{https://ieeexplore.ieee.org/document/9458932/}
\BIBentrySTDinterwordspacing

\bibitem{TraceTogether}
``{{TraceTogether}},'' https://www.tracetogether.gov.sg.

\bibitem{hatkeUsingBluetoothLow2020}
G.~F. Hatke, M.~Montanari, S.~Appadwedula, M.~Wentz, J.~Meklenburg, L.~Ivers,
  J.~Watson, and P.~Fiore, ``Using {{Bluetooth Low Energy}} ({{BLE}}) {{Signal
  Strength Estimation}} to {{Facilitate Contact Tracing}} for {{COVID}}-19,''
  \emph{arXiv:2006.15711 [eess]}, Jul. 2020.

\bibitem{zaimBluetoothLowEnergy2016}
D.~Zaim and M.~Bellafkih, ``Bluetooth {{Low Energy}} ({{BLE}}) based
  geomarketing system,'' in \emph{2016 11th {{International Conference}} on
  {{Intelligent Systems}}: {{Theories}} and {{Applications}} ({{SITA}})}, Oct.
  2016, pp. 1--6.

\bibitem{yilangwuCICOSystemBased2015}
{Yilang Wu}, {Junbo Wang}, {Lei Jing}, {Yinghui Zhou}, and {Zixue Cheng}, ``A
  {{CICO}} system based on {{BLE}} proximity,'' in \emph{2015 {{IEEE}} 7th
  {{International Conference}} on {{Awareness Science}} and {{Technology}}
  ({{iCAST}})}, Sep. 2015, pp. 180--183.

\bibitem{caoDistanceEstimationMethods2018}
Y.~Cao, X.~Lu, Z.~Zhao, X.~Ji, and Y.~Yan, ``Distance {{Estimation Methods}} in
  {{Vehicular Application}}: {{An Experimental Study}},'' in \emph{2018 18th
  {{International Conference}} on {{Control}}, {{Automation}} and {{Systems}}
  ({{ICCAS}})}, Oct. 2018, pp. 1752--1757.

\bibitem{9496688}
G.~Li, S.~Hu, S.~Zhong, W.~L. Tsui, and S.-H.~G. Chan, ``vcontact: Private
  wifi-based iot contact tracing with virus lifespan,'' \emph{IEEE Internet of
  Things Journal}, pp. 1--1, 2021.

\bibitem{SEC14}
S.~He and S.-H.~G. Chan, ``Sectjunction: {Wi}-{Fi} {Indoor} {Localization}
  based on {Junction} of {Signal} {Sectors},'' in \emph{Proceedings of {IEEE}
  {ICC} 2014 - {Mobile} and {Wireless} {Networking} {Symposium} ({ICC}'14
  {MWN})}, Sydney, Australia, Jun. 2014, pp. 2611--2616.

\bibitem{yoonEfficientDistanceEstimation2018}
S.~Yoon, J.~Woo, J.~Cho, and C.~Rahm, ``An {{Efficient Distance Estimation
  Method Based}} on {{Bluetooth Low Energy}} and {{Ultrasound}},'' in
  \emph{2018 {{IEEE International Conference}} on {{Consumer Electronics}} -
  {{Asia}} ({{ICCE}}-{{Asia}})}, Jun. 2018, pp. 206--212.

\bibitem{ramasamy2016lidar}
S.~Ramasamy, R.~Sabatini, A.~Gardi, and J.~Liu, ``Lidar obstacle warning and
  avoidance system for unmanned aerial vehicle sense-and-avoid,''
  \emph{Aerospace Science and Technology}, vol.~55, pp. 344--358, 2016.

\bibitem{xiaoNonLineofSightIdentificationMitigation2015}
Z.~Xiao, H.~Wen, A.~Markham, N.~Trigoni, P.~Blunsom, and J.~Frolik,
  ``Non-{{Line}}-of-{{Sight Identification}} and {{Mitigation Using Received
  Signal Strength}},'' \emph{IEEE Transactions on Wireless Communications},
  vol.~14, no.~3, pp. 1689--1702, Mar. 2015.

\bibitem{zhouLiFiLineOfSightIdentification2014}
Z.~Zhou, Z.~Yang, C.~Wu, W.~Sun, and Y.~Liu, ``{{LiFi}}:
  {{Line}}-{{Of}}-{{Sight}} identification with {{WiFi}},'' in \emph{{{IEEE
  INFOCOM}} 2014 - {{IEEE Conference}} on {{Computer Communications}}}, Apr.
  2014, pp. 2688--2696.

\bibitem{shankarProximitySensingModeling2020a}
S.~Shankar, R.~Kanaparti, A.~Chopra, R.~Sukumaran, P.~Patwa, M.~Kang, A.~Singh,
  K.~P. McPherson, and R.~Raskar, ``Proximity {{Sensing}}: {{Modeling}} and
  {{Understanding Noisy RSSI}}-{{BLE Signals}} and {{Other Mobile Sensor Data}}
  for {{Digital Contact Tracing}},'' \emph{arXiv:2009.04991 [cs, eess]}, Dec.
  2020.

\bibitem{chandelProxiTrakRobustSolution2020}
V.~Chandel, S.~Banerjee, and A.~Ghose, ``\BIBforeignlanguage{en}{{{ProxiTrak}}:
  A robust solution to enforce real-time social distancing \& contact tracing
  in enterprise scenario},'' in \emph{\BIBforeignlanguage{en}{Adjunct
  {{Proceedings}} of the 2020 {{ACM International Joint Conference}} on
  {{Pervasive}} and {{Ubiquitous Computing}} and {{Proceedings}} of the 2020
  {{ACM International Symposium}} on {{Wearable Computers}}}}.\hskip 1em plus
  0.5em minus 0.4em\relax {Virtual Event Mexico}: {ACM}, Sep. 2020, pp.
  503--511.

\bibitem{jimenezFindingObjectsUsing2017a}
A.~R. Jim{\'e}nez and F.~Seco, ``Finding objects using {{UWB}} or {{BLE}}
  localization technology: {{A}} museum-like use case,'' in \emph{2017
  {{International Conference}} on {{Indoor Positioning}} and {{Indoor
  Navigation}} ({{IPIN}})}, Sep. 2017, pp. 1--8.

\bibitem{tc4tlChallenge2020}
\BIBentryALTinterwordspacing
{National Institute of Standards and Technology (NIST) and MIT PACT project},
  ``{TC4TL Challenge},'' 2020. [Online]. Available:
  \url{https://tc4tlchallenge.nist.gov/}
\BIBentrySTDinterwordspacing

\bibitem{ercegEmpiricallyBasedPath1999}
V.~Erceg, L.~J. Greenstein, S.~Y. Tjandra, S.~R. Parkoff, A.~Gupta, B.~Kulic,
  A.~A. Julius, and R.~Bianchi, ``An empirically based path loss model for
  wireless channels in suburban environments,'' \emph{IEEE Journal on Selected
  Areas in Communications}, vol.~17, no.~7, pp. 1205--1211, Jul. 1999.

\bibitem{phillipsSurveyWirelessPath2013}
C.~Phillips, D.~Sicker, and D.~Grunwald, ``A {{Survey}} of {{Wireless Path Loss
  Prediction}} and {{Coverage Mapping Methods}},'' \emph{IEEE Communications
  Surveys Tutorials}, vol.~15, no.~1, pp. 255--270, First 2013.

\bibitem{alqathradyImprovingBLEDistance2017}
M.~Al~Qathrady and A.~Helmy, ``Improving {{BLE Distance Estimation}} and
  {{Classification Using TX Power}} and {{Machine Learning}}: {{A Comparative
  Analysis}},'' in \emph{Proceedings of the 20th {{ACM International
  Conference}} on {{Modelling}}, {{Analysis}} and {{Simulation}} of
  {{Wireless}} and {{Mobile Systems}}}, ser. {{MSWiM}} '17.\hskip 1em plus
  0.5em minus 0.4em\relax {New York, NY, USA}: {Association for Computing
  Machinery}, Nov. 2017, pp. 79--83.

\bibitem{montanariStudyBluetoothLow2017}
A.~Montanari, S.~Nawaz, C.~Mascolo, and K.~Sailer, ``A {{Study}} of {{Bluetooth
  Low Energy}} performance for human proximity detection in the workplace,'' in
  \emph{2017 {{IEEE International Conference}} on {{Pervasive Computing}} and
  {{Communications}} ({{PerCom}})}, Mar. 2017, pp. 90--99.

\bibitem{marateaNonParametricRobust2018}
A.~Maratea, S.~Gaglione, A.~Angrisano, G.~Salvi, and A.~Nunziata, ``Non
  parametric and robust statistics for indoor distance estimation through
  {{BLE}},'' in \emph{2018 {{IEEE International Conference}} on {{Environmental
  Engineering}} ({{EE}})}, Mar. 2018, pp. 1--6.

\bibitem{zafariEnhancingAccuracyIBeacons2017}
F.~Zafari, I.~Papapanagiotou, M.~Devetsikiotis, and T.~J. Hacker, ``Enhancing
  the accuracy of {{iBeacons}} for indoor proximity-based services,'' in
  \emph{2017 {{IEEE International Conference}} on {{Communications}}
  ({{ICC}})}, May 2017, pp. 1--7.

\bibitem{lamImprovedDistanceEstimation2018}
C.~H. Lam, P.~C. Ng, and J.~She, ``Improved {{Distance Estimation}} with {{BLE
  Beacon Using Kalman Filter}} and {{SVM}},'' in \emph{2018 {{IEEE
  International Conference}} on {{Communications}} ({{ICC}})}, May 2018, pp.
  1--6.

\bibitem{pallaviOverviewPracticalAttacks2019}
S.~Pallavi and V.~A. Narayanan, ``An {{Overview}} of {{Practical Attacks}} on
  {{BLE Based IOT Devices}} and {{Their Security}},'' in \emph{2019 5th
  {{International Conference}} on {{Advanced Computing Communication Systems}}
  ({{ICACCS}})}, Mar. 2019, pp. 694--698.

\bibitem{mackeyImprovingBLEBeacon2020}
A.~Mackey, P.~Spachos, L.~Song, and K.~N. Plataniotis, ``Improving {{BLE Beacon
  Proximity Estimation Accuracy Through Bayesian Filtering}},'' \emph{IEEE
  Internet of Things Journal}, vol.~7, no.~4, pp. 3160--3169, Apr. 2020.

\bibitem{ProcedurallyGeneratedEnvironments}
``Procedurally generated environments for simulating {{RSSI}}- localization
  applications | {{Proceedings}} of the 20th {{Communications}} \& {{Networking
  Symposium}},'' https://dl.acm.org/doi/10.5555/3107979.3107986.

\bibitem{ioffeBatchNormalizationAccelerating2015}
S.~Ioffe and C.~Szegedy, ``Batch {{Normalization}}: {{Accelerating Deep Network
  Training}} by {{Reducing Internal Covariate Shift}},'' \emph{arXiv:1502.03167
  [cs]}, Mar. 2015.

\bibitem{misraMishSelfRegularized2020}
D.~Misra, ``Mish: {{A Self Regularized Non}}-{{Monotonic Activation
  Function}},'' \emph{arXiv:1908.08681 [cs, stat]}, Aug. 2020.

\bibitem{apicellaSurveyModernTrainable2021}
A.~Apicella, F.~Donnarumma, F.~Isgr{\`o}, and R.~Prevete, ``A survey on modern
  trainable activation functions,'' \emph{Neural Networks}, vol. 138, pp.
  14--32, Jun. 2021.

\bibitem{Goodfellow-et-al-2016}
I.~Goodfellow, Y.~Bengio, and A.~Courville, \emph{Deep Learning}.\hskip 1em
  plus 0.5em minus 0.4em\relax MIT Press, 2016,
  \url{http://www.deeplearningbook.org}.

\bibitem{chenJointDemosaickingDenoising2021}
J.~Chen, S.~Wen, and S.-H.~G. Chan, ``Joint {{Demosaicking}} and {{Denoising}}
  in the {{Wild}}: The {{Case}} of {{Training Under Ground Truth
  Uncertainty}},'' \emph{arXiv:2101.04442 [cs, eess]}, Jan. 2021.

\bibitem{10.1214/09-SS057}
\BIBentryALTinterwordspacing
J.~Pearl, ``{Causal inference in statistics: An overview},'' \emph{Statistics
  Surveys}, vol.~3, no. none, pp. 96 -- 146, 2009. [Online]. Available:
  \url{https://doi.org/10.1214/09-SS057}
\BIBentrySTDinterwordspacing

\bibitem{huangCorrectingSampleSelection2006b}
J.~Huang, A.~Smola, A.~Gretton, K.~Borgwardt, and B.~Sch{\"o}lkopf,
  ``Correcting {{Sample Selection Bias}} by {{Unlabeled Data}}.'' in
  \emph{Advances in {{Neural Information Processing Systems}} 19:
  {{Proceedings}} of the 2006 {{Conference}}, 601-608 (2007)}, vol.~19, Jan.
  2006, pp. 601--608.

\bibitem{Boyd2006CO}
Boyd, Vandenberghe, and Faybusovich, ``\BIBforeignlanguage{eng}{Convex
  optimization},'' \emph{\BIBforeignlanguage{eng}{IEEE transactions on
  automatic control}}, vol.~51, no.~11, pp. 1859--1859, 2006.

\bibitem{courbariauxBinarizedNeuralNetworks2016}
M.~Courbariaux, I.~Hubara, D.~Soudry, R.~{El-Yaniv}, and Y.~Bengio, ``Binarized
  {{Neural Networks}}: {{Training Deep Neural Networks}} with {{Weights}} and
  {{Activations Constrained}} to +1 or -1,'' \emph{arXiv:1602.02830 [cs]}, Mar.
  2016.

\end{thebibliography}
\end{document}